\documentclass[11pt,letterpaper]{article}

\usepackage{cite}
\usepackage{amsmath,amssymb,amsfonts}
\usepackage{algorithmic}
\usepackage{graphicx}
\usepackage{textcomp}
\usepackage{xcolor}

\usepackage[letterpaper,margin=1.0in]{geometry}
\usepackage{url}
\usepackage{amssymb}
\usepackage{graphicx}
\usepackage{amsmath}
\usepackage{amsfonts}
\usepackage{amssymb}
\usepackage{amsthm}
\usepackage{algorithm}
\usepackage{algorithmic}
\usepackage[pdftex,
            colorlinks,
            plainpages=false,
            hyperindex,
            bookmarksopen,
            linkcolor=black,
            citecolor=black,
            urlcolor=blue]{hyperref}
\usepackage[labelfont=bf]{subcaption}
\usepackage{multirow}
\usepackage{xcolor}
\usepackage{float}
\usepackage{authblk}
\usepackage{braket}
\usepackage{mathtools}
\usepackage[normalem]{ulem}

\newcommand{\vect}[1]{\boldsymbol{#1}}

\title{Classical and Quantum Solvers for Joint Network/Servers Power Optimization}

\author[1,2]{Davide Ferrari}
\author[1,2]{Michele Amoretti}
\author[3]{A. Manzalini}
\affil[1]{Department of Engineering and Architecture - University of Parma, Italy}
\affil[2]{Quantum Information Science @ University of Parma, Italy}
\affil[3]{Department of Innovation of TIM, 10148 Turin, Italy}

\date{}

\begin{document}

\maketitle

\begin{abstract}
The digital transformation that Telecommunications and ICT domains are crossing today, is posing several new challenges to Telecom Operators. These challenges require solving complex problems such as: dimensioning and scheduling of virtual/real resources in data centers; automating real-time management/control and orchestration of networks processes; optimizing energy consumption; and overall, ensuring networks and services stability. These problems are usually tackled with methods and algorithms that find suboptimal solutions, for computational efficiency reasons.
In this work, we consider a Virtual Data Center scenario where virtual machine consolidation must be performed with joint minimization of network/servers power consumption. For this scenario, we provide an ILP model, the equivalent binary model and the steps towards the equivalent Quadratic Unconstrained Binary Optimization (QUBO) model that is suitable for being solved by means of quantum optimization algorithms. Finally, we compare the computational complexity of classical and quantum solvers from a theoretical perspective.\\
\textbf{Keywords:} Virtual Data Center, joint network/servers power optimization, QUBO models, quantum solvers
\end{abstract}

\section{Introduction}
\label{sec:introduction}

Telecommunications and ICT domains are crossing, today, a profound  Digital Transformation. Major drivers include pervasive ultra-broadband fixed-mobile connectivity with very low latency (e.g., 5G \cite{Soldani2015}) and the increased flexibility and programmability of networks, thanks to Software Defined Networking (SDN), Network Function Virtualization (NFV) \cite{Clayman2014}, and a deeper integration of Artificial Intelligence with network and service platforms (e.g., Cloud and Edge Computing). Future network and service infrastructures, such as 6G \cite{Letaief2019}, will face data traffic growth challenges in the context of advanced service and application scenarios, such as ultra-massive scale communications for ambient intelligence, holographic telepresence, tactile Internet, new paradigms for brain computer interactions, innovative forms of communications.

This Digital Transformation is posing several new challenges to Telecom Operators, which require solving complex problems such as: dimensioning and scheduling of virtual/real resources in Virtual Data Centers (VDCs); automating real-time management/control and orchestration of networks processes; predicting and reacting to epidemic spreading of software viruses and security attacks; optimizing energy consumption; and overall, ensuring networks and services stability. Most of these challenges imply solving optimization problems with multiple constraints. Today these problems are tackled with methods and algorithms that find suboptimal solutions, because of the excessive cost of finding an optimal solution. Specifically, the evolution towards the model of Cloud-native distributed networking and execution environments is putting under the spotlight the power optimization problem \cite{ZhengJointPower2014b,Lu2018,Jayanetti2019}.

The main objective of this work is to compare classical and quantum solvers for the joint network/servers power optimization problem, in terms of computational complexity. To this purpose, we consider a VDC scenario where virtual machine (VM) consolidation must be performed with joint minimization of network/servers power consumption. For this scenario, we provide an ILP model, the equivalent binary model and the steps towards the equivalent Quadratic Unconstrained Binary Optimization (QUBO) model that is suitable for being solved by means of quantum optimization algorithms. 
Finally, we compare the computational complexity of classical and quantum solvers when they are used to find a solution to the considered optimization problem with the proposed models (respectively, binary and QUBO). 

By means of theoretical analysis and experiments, we show that classical solvers are not space and time efficient. By means of theoretical considerations, we show that, conversely, quantum solvers are space and time efficient in particular, the required quantum memory is polynomial in the problem size), but the number or required qubits is still too large for current quantum devices.

The paper is organized as follows. In Section \ref{sec:models}, we illustrate considered system architecture and the problem formulation models. In Section \ref{sec:analysis}, we discuss the performance of classical and quantum solvers, with respect to the proposed models. Finally, in Section \ref{sec:conclusions}, we draw some conclusions.


\section{Related Work}
\label{sec:related}

Energy efficiency is an important and fundamental research issue in Cloud Computing. The energy efficiency of VDCs using SDN technology is improved by reducing the energy consumption in VM and network using placement, consolidation, and overbooking techniques \cite{Son2018}.

Joint power optimization by utilizing VM placement and flow routing is a kind of \textit{VDC embedding problem}, i.e., a problem about finding the optimal mappings between virtual resource to physical resource, which is NP-Hard (indeed, it is a multidimensional bin packing problem \cite{Christensen2017}). 
Jin et al. \cite{Jin2013} formulated the joint power optimization problem as an integer linear program (ILP), which is not a practical solution due to high complexity. To practically and effectively combine host and network based optimization, the authors proposed a unified representation method that converts the VM placement problem to a routing problem. In addition, to accelerate processing the large number of servers and an even larger number of VMs, they described a parallelization approach that divides the network into clusters for parallel processing. Furthermore, to quickly find efficient paths for flows, the authors proposed a fast topology oriented multipath routing algorithm that uses depth-first search to quickly traverse between hierarchical switch layers and uses the best-fit criterion to maximize flow consolidation.

Han et al. \cite{Han2015} considered the VDC embedding problem in multiple physical data centers and proposed an SDN Assisted VDC Embedding solution (SAVE), including two VDC embedding algorithms (denoted as Resource Raising and Resource Falling, respectively) and a Traffic Engineering (TE) algorithm to realize the network fabrics consolidation. All the proposed algorithms are greedy bin packing heuristics which quickly produce a near-optimal solution.

In the recent literature \cite{Carpio2017,Sun2017,Guo2020,Pham2020,Wang2021}, further heuristic algorithms have been proposed to tackle the VDC embedding problem, always starting from an ILP optimization model. To the best of our knowledge, no one attempted to formulate the problem using a QUBO model to be solved using a quantum computer. In the remainder of the paper, we fill this gap and we provide a theoretical analysis of classical and quantum solvers.  

\begin{figure}
    \centering
    \includegraphics[width=\columnwidth]{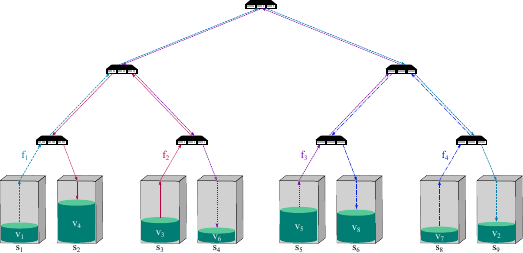}
    \caption{System of servers connected through switches that are organized in a tree topology.}
    \label{fig:server_topology}
\end{figure}

\section{Models}
\label{sec:models}

We assume a system of physical servers connected through switches that are organized in a tree topology. Each server may host several VMs. Data are transmitted into flows between VMs, through links that connect server and switches. An example instance is illustrated in Figure~\ref{fig:server_topology}, in which 8 servers host 8 virtual machines and data flows between couples of VMs may need to traverse multiple switches.    

\begin{table}
    \centering
    \small
    \begin{tabular}{|c|c|}
        \hline
        \textbf{Notation} & \textbf{Meaning} \\
        \hline
        $v_{ji}$ & the $j$th VM on the $i$th server \\
        \hline
        $u(v_{ji})$ & the normalized CPU utilization of $v_{ji}$ \\
        \hline
        $P_{i}^{idle}$ & the idle power consumption of node $i$ \\
        \hline
        $P_{i}^{dyn}$ & the maximum dynamic power of node $i$ \\
        \hline
        $P_{i}^{s}$ & the total power consumption of server $i$ \\
        \hline
        $P_{k}^{sw}$ & the total power consumption of switch $k$ \\
        \hline
        $w_i$ & the number of VMs assigned to server $i$ \\
        \hline
        $P_{k}^{idle}$ & the idle power consumption of switch $k$ \\
        \hline
        $P_{rk}$ & the power consumption of port $r$ on switch $k$ \\
        \hline
        $r_k$ & the number of active ports on switch $k$  \\
        \hline
        $N$ & the total number of VMs \\
        \hline
        $M$ & the total number of servers \\
        \hline
        $K$ & the total number of switches \\
        \hline
        $L$ & the total number of links \\
        \hline
        $F$ & the total number of flows \\
        \hline
        $F_l$ & the total number of flows on link $l$ \\
        \hline
        $d_{f,l}$  & the data rate of flow $f$ on link $l$ \\
        \hline
        $C_l$ & the capacity of link $l$ \\
        \hline
        $C_s$ & the capacity of server $s$ \\
        \hline
        $\tilde{C}_{\{l,s\}}$ & $\lceil\log_2 C_{\{l,s\}}\rceil$\\
        \hline
    \end{tabular}
    \caption{Notation for the problem formulation}
    \label{tab:notation}
\end{table}

\subsection{ILP Model}
\label{sec:preliminary_model}

Using the notation summarized in Table~\ref{tab:notation}, we formulate the joint network/servers power optimization problem using an ILP model that extends and further refines those proposed by Jin et al. \cite{Jin2013} and Zheng et al. \cite{ZhengJointPower2014b} (Eq. 1 to 6).
\begin{align}
    \min &\left(\sum_{i=1}^{M} P_{i}^{s} + \sum_{k=1}^{K} P_{k}^{sw}\right) \\
    \text{s.t.} \; &\sum_{i=1}^{M} w_i = N \\
    &\sum_{j=1}^{w_i} u(v_{ji}) \leq 1, \quad \forall i \in \{1,..,M\}\\
    &\sum_{f=1}^{F} d_{f,l} \leq C^l, \quad \forall l \in \{1,..,L\} \label{eq:eq4}
\end{align}
with:
\begin{align}
    P_{i}^{s} &= \begin{cases} P_{i}^{idle}+\smashoperator{\sum_{j=1}^{w_i}} u(v_{ji})P_{i}^{dyn}, & w_i > 0 \\ 0, & w_i = 0 \end{cases}\\
    P_{k}^{sw} &= \begin{cases} P_{k}^{idle}+\smashoperator{\sum_{r=1}^{a_k}} P_{rk}, & \mbox{if switch }k\mbox{ is on} \\ 0, & \mbox{otherwise} \end{cases}
\end{align}

Having $a_k$ active ports on a switch means that the sum of input and output flows that cross the switch is $a_k$. 
Each flow $f$ is identified by a VM pair: $src(f)$ and $dst(f)$. In the above model, a constraint stating that all the flows are allocated (by means of the VM-server allocation and by the set of active switches) is missing. Such a constraint should be like the one of Eq. (2), which states that all the VMs are allocated to some servers.

To solve this issue, the model must be refined. Therefore, $\rho$ is defined such that  $\rho(f,(n_1,n_2))=1$ if part of the $f$th flow goes from $n_1$ to $n_2$, and $\rho(f,(n_1,n_2))=0$ otherwise, where $n_1$ and $n_2$ are adjacent nodes in $\mathcal{M}\cup \mathcal{K}$, being $\mathcal{M}$ the set of servers and $\mathcal{K}$ the set of switches.
The definition of $\rho$ allows us to add the following constraint:
\begin{equation}
    v_{src(f),i} - v_{dst(f),i} = \sum_{k=1}^{K} \rho(f,(i,k)) - \sum_{k=1}^{K} \rho(f,(k,i)),
\end{equation}
for each flow $f$ and for each server $i$. This constraint forces the allocation of all the $F$ flows, whose source and destination may be hosted by the same server or by different servers.

Another constraint that must be introduced to refine the model is
\begin{align}
\sum_{n \in \mathcal{M}\cup \mathcal{K}} \rho(f,(n,k)) = \sum_{n \in \mathcal{M}\cup \mathcal{K}} \rho(f,(k,n)),
\end{align}
for each flow $f$ and for each switch $k$. This constraint ensures that flows are assigned to complete paths, from source to destination.

Furthermore, Eq.~\ref{eq:eq4} can be restated as
\begin{small}
\begin{equation}
    \sum_{f=1}^{F}d_{f,l_{n_1,n_2}}  P(f_{n},(n_1,n_2)) \leq C_{l_{n_1,n_2}} on(n_1)on(n_2),\, \forall (n_1,n_2),
    \label{eq:quadratic_capacity}
\end{equation}
\end{small}
where $P(f_{n},(n_1,n_2))=[\rho(f_{n},(n_1,n_2))+\rho(f_{n},(n_2,n_1))]$, $n_1$ and $n_2$ are adjacent nodes (each of them being either a server or a switch), $l_{n_1n_2}$ is a link between them and $on(n) = 1$ if node $n$ is on, $0$ otherwise.

\subsection{Binary Model}
\label{sec:binary_model}

The binary version of the model presented in Section \ref{sec:preliminary_model} is presented below.
\begin{small}
\begin{align}
    \nonumber
    \min \:&\sum_{i=1}^{M}\left(s_i P_{i}^{idle} + P_{i}^{dyn}\sum_{j=1}^{N} u(v_{ji})v_{ji} \right) +\\&+ \sum_{k=1}^{K}\left(sw_k P_{k}^{idle}+\sum_{r=1}^{a_k} P_{rk} \right)
\end{align}
\end{small}
s.t.    
\begin{small}
\begin{align}
    &\sum_{j=1}^{N} u(v_{ji})v_{ji} \leq C_{s}s_i, \quad \forall i \in \{1,..,M\} 
    \label{eq:server_cap}\\
    &\sum_{i=1}^{M} v_{ji} = 1, \quad \forall j \in \{1,..,N\} \label{eq:c2}\\
    \label{eq:source}
    &\sum_{k=1}^{K} \rho(f, (i,k)) \leq v_{src(f), i}, \quad \forall i \in \{1,..,M\}, \forall f \in \{1,..,F\}\\
    \label{eq:dest}
    &\sum_{k=1}^{K} \rho(f, (k,i)) \leq v_{dst(f), i}, \quad \forall i \in \{1,..,M\}, \forall f \in \{1,..,F\}\\
    \nonumber
    &v_{src(f),i} - v_{dst(f),i} = \sum_{k=1}^{K} \rho(f,(i,k)) - \sum_{k=1}^{K} \rho(f,(k,i)), \\&\hspace{3.5cm}\forall i \in \{1,..,M\}, \forall f \in \{1,..,F\}\\
    \nonumber
    &\sum_{n \in \mathcal{M}\cup \mathcal{K}}^{}\rho(f,(n,k)) = \sum_{n \in \mathcal{M}\cup \mathcal{K}}^{}\rho(f,(k,n)), \\&\hspace{3.5cm} \forall f \in \{1,..,F\}, \forall k \in \{1,..,K\}\\
    \label{eq:capcity_bin}
    &\sum_{f=1}^{F}d_{f,l_{n_1,n_2}} P(f_{n},(n_1,n_2)) \leq C_{l_{n_1n_2}} on(n_1, n_2), \quad \forall l_{n_1,n_2}\\
    \label{eq:link_on_1}
    &on(n_1, n_2) \leq on(n_1), \quad \forall l_{n_1,n_2}\\
    \label{eq:link_on_2}
    &on(n_1, n_2) \leq on(n_2), \quad \forall l_{n_1,n_2}
\end{align}
\end{small}

Here, the following binary variables are used:
\begin{itemize}
    \item $s_i$ is a binary variable for $i$th server and $s_i = 1$ means that server $i$ is on;
    \item $sw_k$ is a binary variable for $k$th switch and $sw_k = 1$ means that switch $k$ is on;
\end{itemize}

Eq.~\ref{eq:server_cap} ensures that no server is overloaded while also enforcing that VMs cannot be assigned to inactive servers. To simplify the conversion to a QUBO problem, we assume that $u(v_{ji}) \in [1,...,10]$. Eq.~\ref{eq:c2} checks that each VM is assigned to exactly one server.

Moreover, quadratic constraints cannot be converted into QUBO quadratic penalty terms, thus we need to add a binary variable for every link and split Eq.~\ref{eq:quadratic_capacity} into Eq.~\ref{eq:capcity_bin}, \ref{eq:link_on_1} and \ref{eq:link_on_2}. There, we have that $on(n_1, n_2) = 1$ if $l_{n_1,n_2}$ is on, $0$ otherwise. Equation~\ref{eq:link_on_1} and Equation~\ref{eq:link_on_2} ensure that a link is on if and only if both endpoints nodes of that link are on.

The total number of binary variables is then
\begin{equation}
    Q = M+MN = M(N+1)+K+2LF+L.
\end{equation}

\subsection{QUBO Model}
\label{sec:qubo_model}

Quantum optimization algorithms can only be applied to QUBO models. In order to be able to convert the model presented in the previous section into a QUBO model, one must first turn all inequality constraints to equality ones. This is done by adding slack variables.

By adding $\tilde{C_s}$ binary slack variables, Eq.~\ref{eq:server_cap} becomes:
\begin{equation}
    \sum_{j=1}^{N} u(v_{ji})v_{ji} = C_s s_i - \smashoperator{\sum_{c=1}^{\tilde{C_s}}} 2^{c-1}slack_{ic}, \quad \forall i \in \{1,..,M\}\\
\end{equation}

Eq.~\ref{eq:source}~and~\ref{eq:dest} need one slack binary variable each, resulting in:
\begin{align}
\nonumber
    &\sum_{k=1}^{K} \rho(f, (i,k)) = v_{src(f), i} - slack_{src(f),i}, \\&\hspace{3.5cm} \forall i \in \{1,..,M\}, \forall f \in \{1,..,F\}\\
    \nonumber
    &\sum_{k=1}^{K} \rho(f, (k,i)) = v_{dst(f), i} - slack_{dst(f),i}, \\&\hspace{3.5cm} \forall i \in \{1,..,M\}, \forall f \in \{1,..,F\}
\end{align}

Eq.~\ref{eq:capcity_bin}, \ref{eq:link_on_1}, and \ref{eq:link_on_2} also need to be converted to equality constraints, becoming:
\begin{align}
    \nonumber
    &\sum_{f=1}^{F}d_{f,l_{n_1,n_2}}  P(f_{n},(n_1,n_2)) = C_{l_{n_1n_2}} on(n_1, n_2) - \\&\hspace{3cm}+\smashoperator{\sum_{c=1}^{\lceil\log_2 C_{l_{n_1,n_2}}\rceil}} 2^{c-1}slack_{l_{n_1,n_2}c}, \quad \forall l_{n_1,n_2}
\end{align}


Lastly, one need $L$ binary variables for the links, and $L\lceil\log_{2} C_{l}\rceil=L\tilde{C_l}$ slack variables for the capacity constraints. Eq.~\ref{eq:link_on_1} and~\ref{eq:link_on_2} fall into a special family of constraints that do not require slack variables to construct penalty terms~\cite{Glover2018}.

The total number of binary variables turns out to be
\begin{equation}
    Q = M(N+\tilde{C_s}+2F+1)+K+2LF+(1+\tilde{C_l})L \\
    \label{eq:q1}.
\end{equation}

Assuming a binary tree network topology, one requires $L=\sum_{t=0}^{\lceil\log_{2} M\rceil-1}\left\lceil\frac{M}{2^{t}}\right\rceil<2M$ variables for the links.
In the worst case scenario, where each link at the lowest level of the tree is full-loaded, a binary tree network would be badly overloaded. To overcome this issue, let us assume instead a \textit{fat tree} topology, where the links capacity is doubled at each higher level. 
There are $\left\lceil\frac{M}{2^t}\right\rceil$ links at each tree level $t \in [0,...,\lceil\log_{2}M\rceil)$, where $t=0$ holds for the lowest level links. The capacity of each link at level $t$ is $C_{l,t}=C_{l}\cdot 2^t$, requiring $\lceil\log_{2}(C_{l}\cdot  2^{t})\rceil=t+\tilde{C_l}$ slack variables for the capacity constraint of each link.
Therefore, with $L~=~\sum_{t=0}^{\lceil\log_{2}M\rceil-1}\left\lceil\frac{M}{2^t}\right\rceil$, Eq.~\ref{eq:q1} becomes:

\begin{equation}
    Q = M(N+\lceil\log_2 C_{s}\rceil+1)+K+2LF+L+ \smashoperator{\sum_{t=0}^{\lceil\log_{2}M\rceil-1}}(t+\tilde{C_l})\left\lceil\frac{M}{2^t}\right\rceil\label{eq:q2}
\end{equation}

Considering that $M$ and $N$ are the dominant variables, we can observe that the QUBO model requires a quantum memory with size $Q = \text{O}(MN)$.

Once all constraints have been transformed, they must be eliminated and added to the objective function as quadratic penalty terms~\cite{Fiacco1990}.

\section{Computational Complexity Analysis}
\label{sec:analysis}

In this section, we discuss the performance of classical and quantum solvers with respect to the proposed binary and QUBO models, respectively.

\subsection{Classical Solvers}

It is well known that 1-0 integer programming is NP-complete \cite{Karp1972}. Approximate solutions can be obtained by means of the \textit{branch-and-bound} (B\&B) framework  \cite{Land1960}.  
To solve a problem $\mathcal{P}$, B\&B iteratively builds a search tree $\mathcal{T}$ of subproblems. A feasible solution (called the incumbent solution) is stored globally. At each iteration, the algorithm selects a new subproblem $\mathcal{S}$ from a list of unexplored subproblems. If no solution can be found with a better objective value than the incumbent one, $\mathcal{S}$ is pruned. If $\mathcal{S}$ cannot be pruned, $\mathcal{S}$ is partitioned into an exhaustive set of subproblems which are then inserted into $\mathcal{T}$ and the incumbent solution is updated. Once no unexplored subproblems remain, the best incumbent solution is returned. To use the B\&B method with an integer programming problem, it is necessary to first solve the problem as a regular linear programming problem without integer restrictions. The optimal solution to the linear programming problem may be integer or not. If it is not an integer, it is used to define two subproblems of the integer programming problem. The relaxed linear programming models corresponding to these subproblems are then solved. If a feasible integer solution does not emerge, the B\&B process continues until a satisfactory approximate solution is found.

Any B\&B algorithm operates in $O(\tau b^d)$ worst-case execution time \cite{Morrison2016}, where $\tau$ is a bound on the length of time needed to explore a subproblem; $b$ is the branching factor of the tree, which is the maximum number of children generated at any node in the tree; and $d$ is the search depth of the tree, which is the length of the longest path from the root of $\mathcal{T}$ to a leaf.  However, the algorithm performance can be substantially improved by the presence of pruning rules. One difficulty is that actual algorithm performance can be highly sensitive to initial conditions. For these reasons, researchers in combinatorial optimization tend to favor experimental evaluations over theoretical asymptotic time bounds, because the latter are hard to come by and usually unduly pessimistic.

\begin{figure}[ht!]
\centering
\begin{tabular}{ cc }
    \multicolumn{2}{c}{
    \begin{minipage}{0.3\textwidth}
    	\centering
        \includegraphics[width=\textwidth]{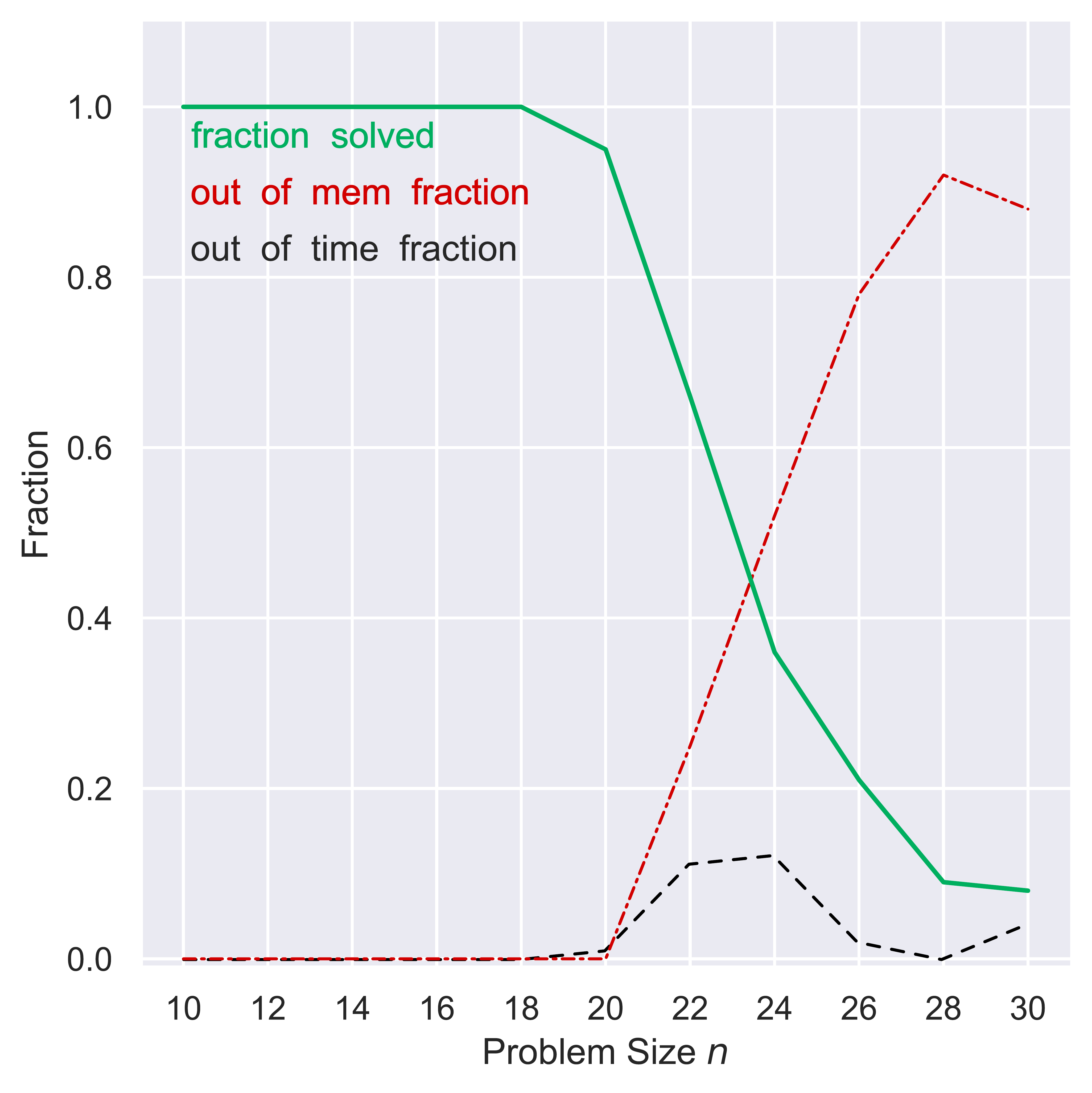}
        \subcaption{}
        \label{fig:fraction}
    \end{minipage}}\\
    \begin{minipage}{0.3\textwidth}
    	\centering
        \includegraphics[width=\textwidth]{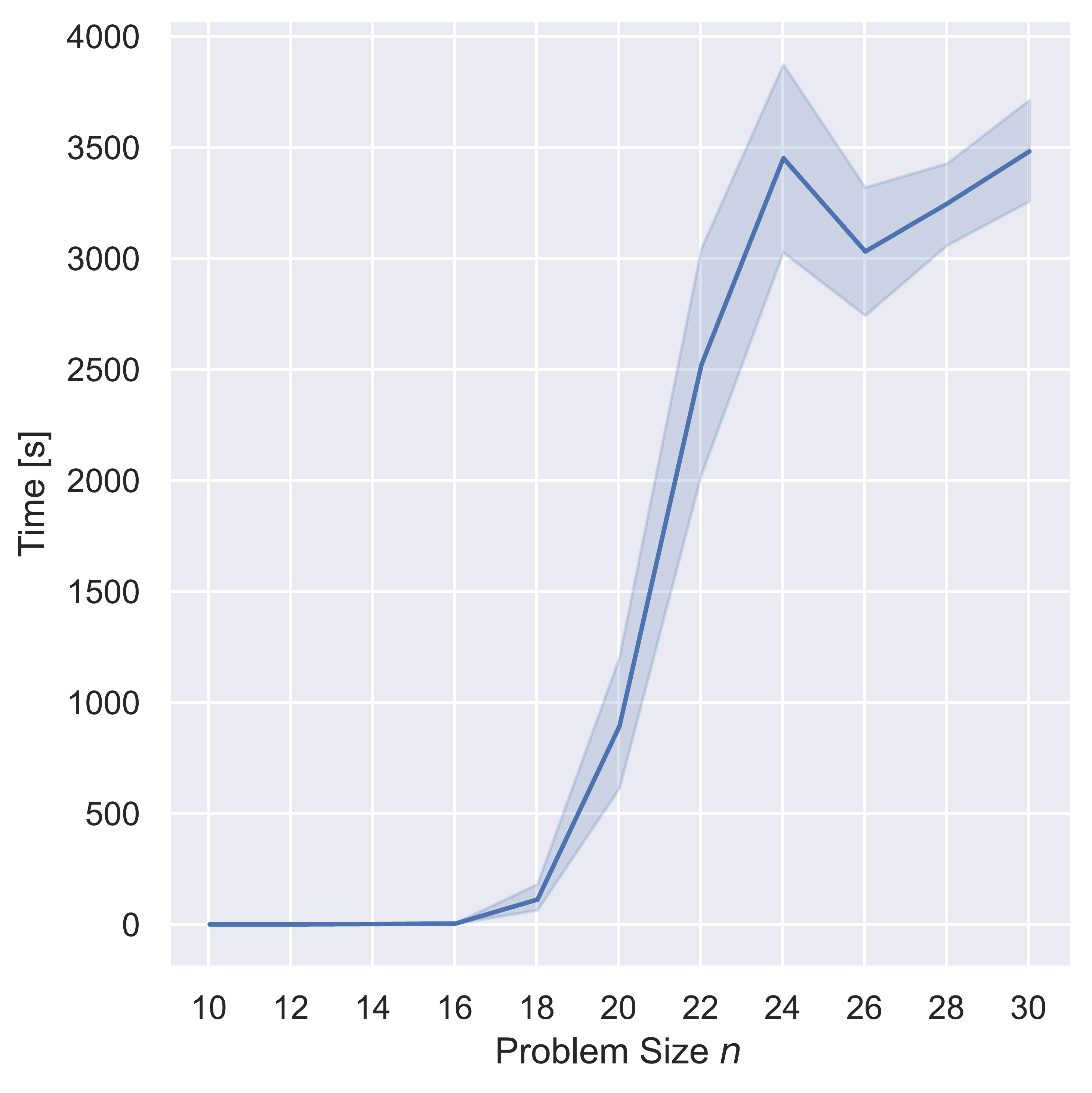}
        \subcaption{}
        \label{fig:time}
    \end{minipage}
    &
    \begin{minipage}{0.3\textwidth}
    	\centering
        \includegraphics[width=\textwidth]{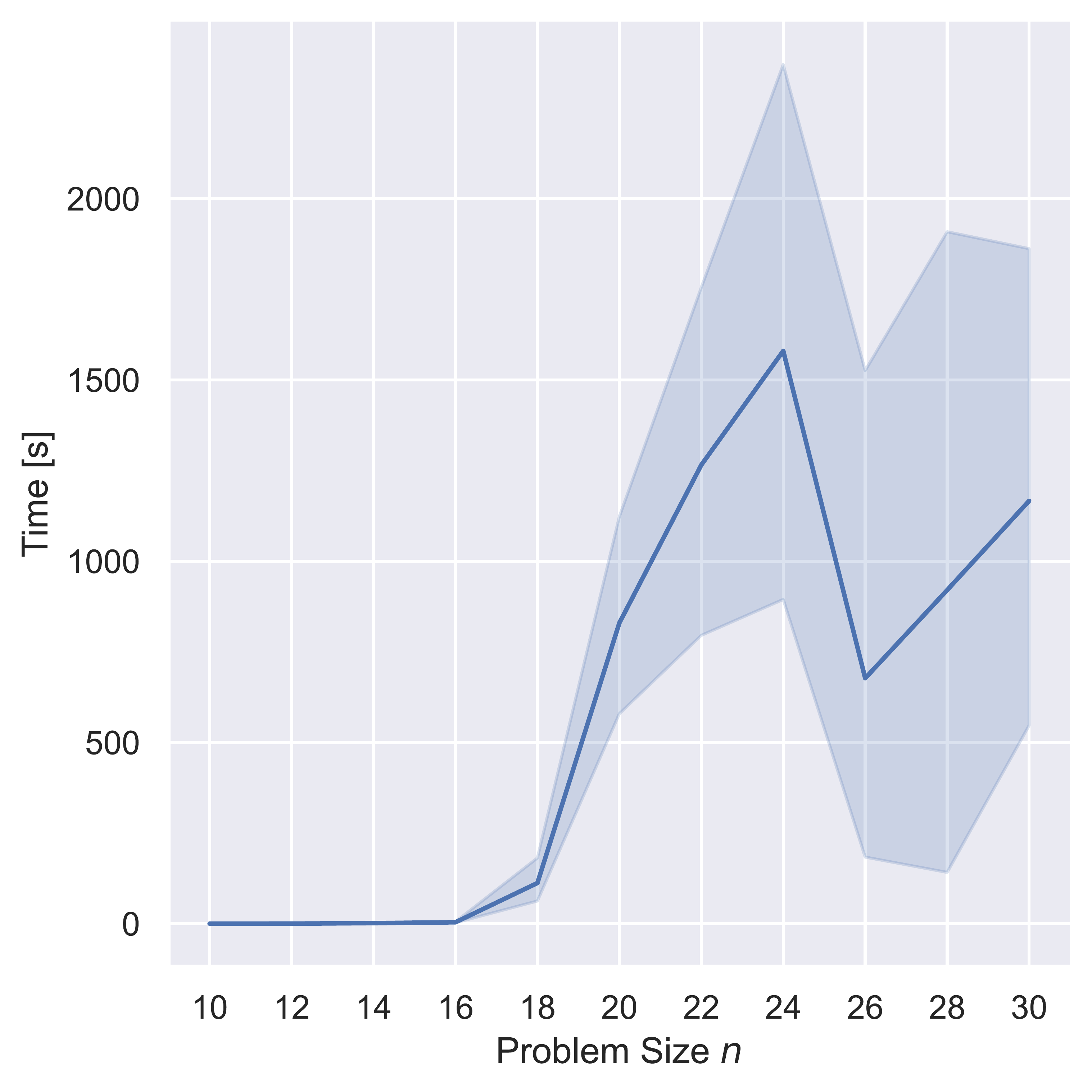}
        \subcaption{}
        \label{fig:solved_time}
    \end{minipage}
    \\
    \begin{minipage}{0.3\textwidth}
    	\centering
        \includegraphics[width=\textwidth]{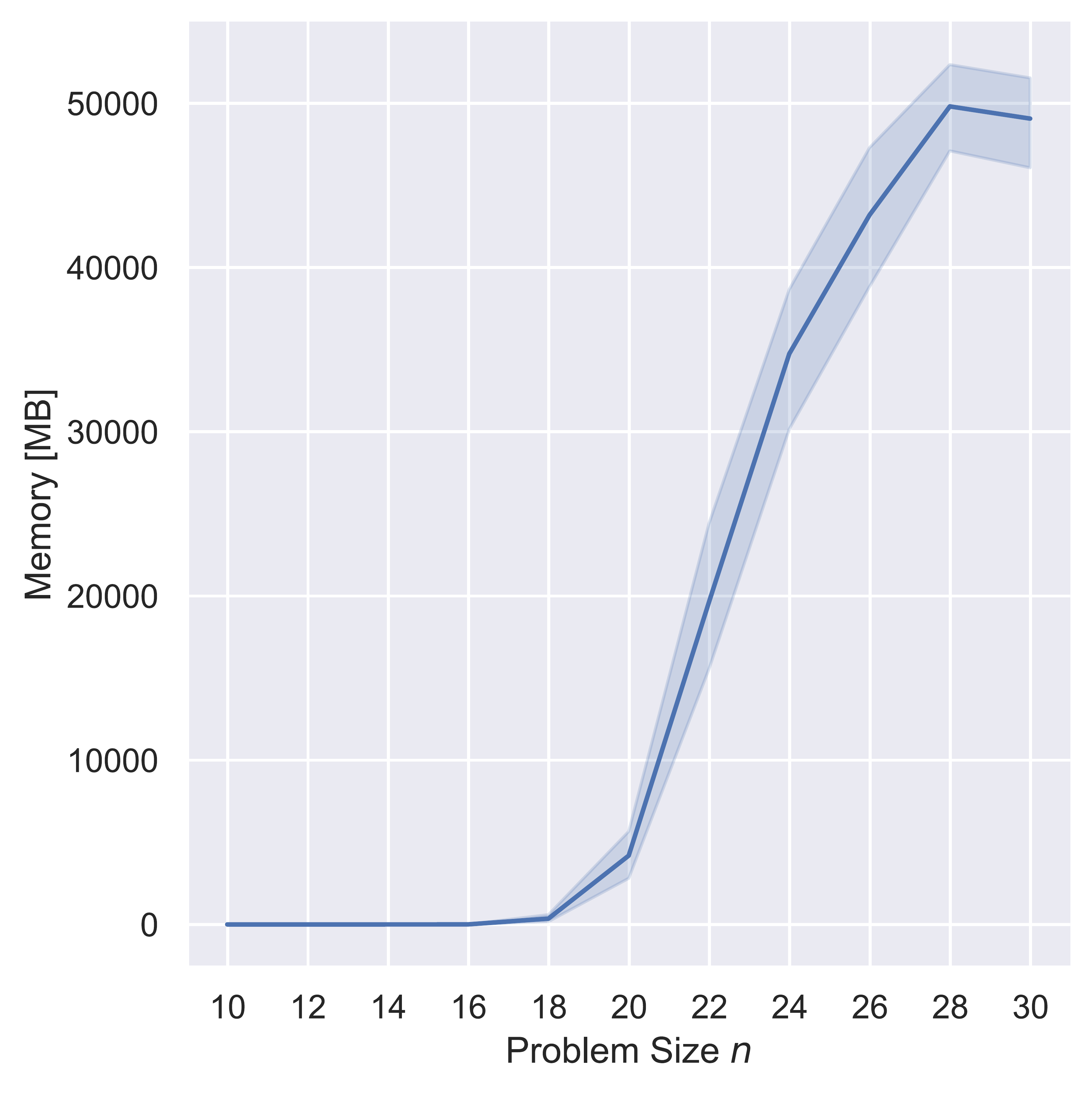}
        \subcaption{}
        \label{fig:mem_usage}
    \end{minipage}
    &
    \begin{minipage}{0.3\textwidth}
		\centering
        \includegraphics[width=\textwidth]{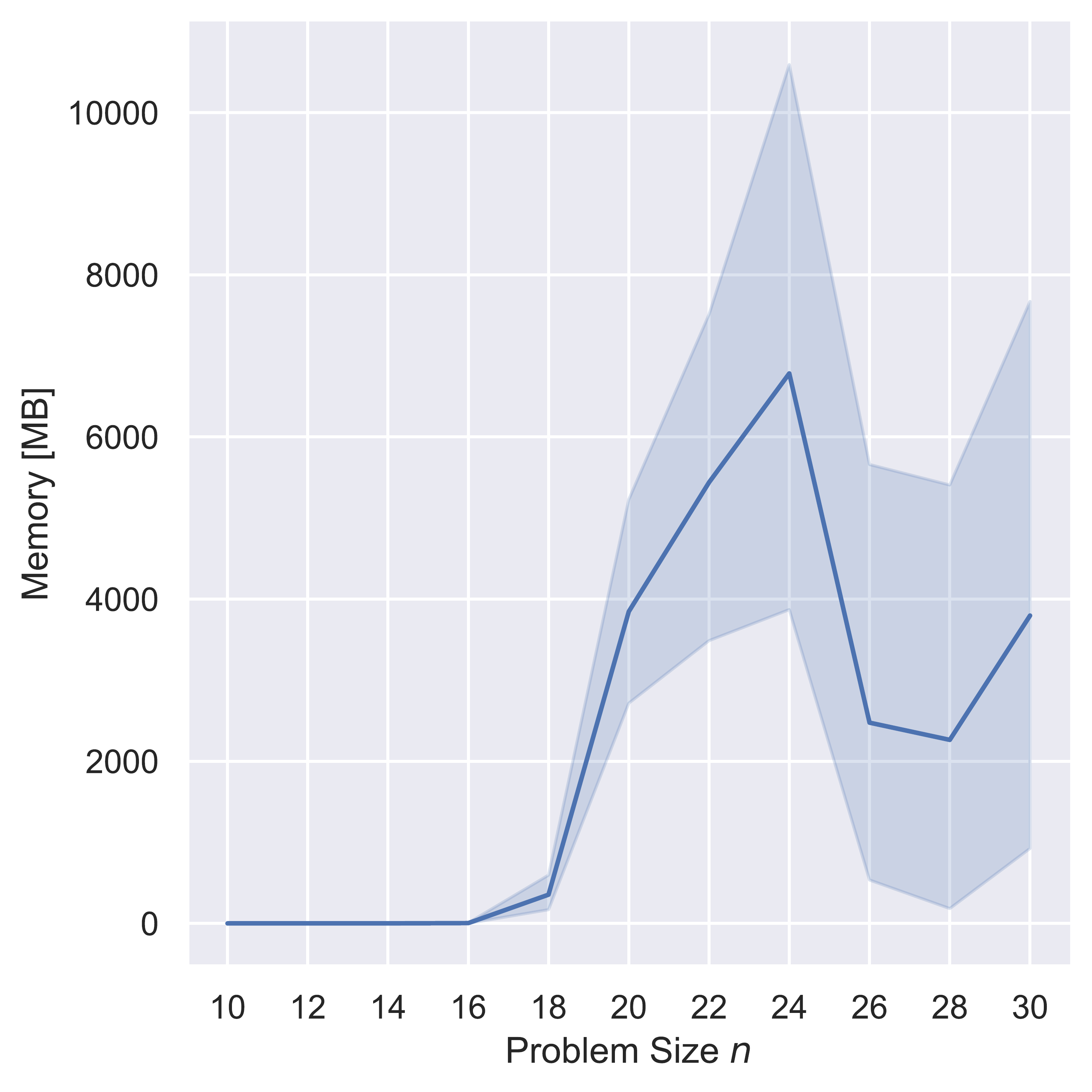}
        \subcaption{}
        \label{fig:solved_mem_usage}
	\end{minipage}
\end{tabular}
\caption{(\ref{fig:fraction}) Fraction of solved, out of time and out of memory instances. (\ref{fig:time}) Solver's time for all instances.
(\ref{fig:solved_time}) Time spent to solve instances, restricted to those that were actually solved.
(\ref{fig:mem_usage}) Solver's memory usage for all instances. (\ref{fig:solved_mem_usage}) Amount of memory used to solve instances, restricted to those that were actually solved.}
\label{fig:performance}
\end{figure}

Many sound software packages exist for solving integer programs using B\&B techniques, including CPLEX \cite{CPLEX}, SYMPHONY \cite{Symphony}, Gurobi \cite{Gurobi}, Cbc \cite{Cbc}. CPLEX has many parameters that allow users to customize the way the CPLEX B\&B algorithm operates. While this variety of parameters provides many different ways to improve performance, one cannot realistically experiment with all the possible combinations of parameter settings.

To solve problem instances based on the binary model that we defined in Section \ref{sec:binary_model}, we used DOcplex Mathematical Programming Modeling, which is CPLEX Python API, on a machine equipped with INTEL XEON E5, reserving 12 cores and 50 GB of RAM. We adopted the Barrier algorithm for the initial relaxation, and we set parallel optimization mode to opportunistic in order to use all available parallelism. Finally, we set a time limit of 2 hours for each solver execution. The considered problem instances consist of $M=n$ servers, $N=n$ VMs, $C_s = C_l = 10$, and $F=n/2$ flows between different VM pairs. Servers and switches have an idle and dynamic power consumption equal to 10 and 1, respectively. VM usage and flow demands have been taken from a normal distribution with mean 4 and then rounded to integers values. The obtained results (Figure~\ref{fig:performance}) show that the solver goes out of time and out of memory for problem instances with size $n > 18$. With $n < 18$, the resource usage grows fast.

\subsection{Quantum Solvers}
In Section \ref{sec:qubo_model}, we have shown that the quantum memory size required by the QUBO model is $Q = O(MN)$, i.e., polynomial in the total number of available servers multiplied by the total number of VMs to be allocated. The next question is: what is the best execution time that can be achieved by means of a quantum solver? To answer this question, we need to find the quantum optimization circuit with the lowest depth, i.e., the one corresponding to the computation with the lowest number of sequential steps. 

Grover's search algorithm \cite{Grover1996} can be used inside a global optimization algorithm \cite{Baritompa2005}. The process starts with an initial solution (either random or incumbent). Grover's algorithm is used to find any solution that has a lower cost than the initial one. In case of success, the initial solution is replaced with the new one. The procedure is iterated until some criteria are met. In case of failure, the Grover’s search block is updated before a new iteration is executed. The depth of the corresponding quantum circuit and the subroutine for updating such a circuit are both  polynomial in the number of variables (qubits) $Q$. The number of iterations needed to find the solution is O($\sqrt{2^Q}$). If the QUBO problem has $t$ solutions, the number of iterations reduces to O($\sqrt{2^Q/t}$). 

Another hybrid quantum/classical approach is to map the QUBO problem to an Ising Hamiltonian $\hat{H}$ and find its ground state using variational algorithms like Variational Quantum Eigensolver (VQE) \cite{Peruzzo2014} or Quantum Approximate Optimization Algorithm (QAOA) \cite{FarhiQuantumApproximate2014}.

Let us consider an Ising spin glass model with Hamiltonian constituted as a summation of Pauli-$Z$ operators (the two eigenvalues $\pm 1$ correspond to the positive and negative spin). Finding the ground state for an Ising Hamiltonian can be seen as a quadratic unconstrained binary optimization problem, and it inherits its hardness. In general, computing the partition function of an Ising model is NP-complete \cite{Barahona1982}. Therefore, any problem in NP can be translated to an Ising model. 

Combinatorial problems have a natural mapping to Ising spin glass models. Consider a quadratic unconstrained binary optimization (QUBO) problem:
\begin{align}
\label{eq:quadmodel}
\min &\left(c ^Tx + x^TQx \right) \\
\text{s.t. } & x \in \{0,1\}^n, c \in \mathcal{R}^n, Q \in \mathcal{R}^{n \times n},
\end{align}
then transform it into an Ising model using the substitutions \begin{equation}
x_j = \frac{y_j^Z +1}{2}, 
\end{equation}
where $x_j \in \{0,1\}$ and $y_j^Z \in \{-1,1\}$ for $j=1,..,n$. The superscript $Z$ is used to distinguish $\pm 1$ spins from $0-1$ variables. Since (\ref{eq:quadmodel}) is a quadratic model, the substitution yields a  summation of terms, each of which contains one or two $y_j^Z$ variables. The Hamiltonian is then a summation of weighted tensor products of Pauli-$Z$ operators, where each term of the summation contains at most two $Z$s. Furthermore, since $Z$ is diagonal, the resulting Hamiltonian is diagonal. If the original binary quadratic optimization problem is constrained, the approach mentioned above can still be applied by adding appropriate (quadratic) penalties for constraint violations in the objective function \cite{Nannicini2019}. 

A variational quantum circuit can be used to prepare an Ans\"atz $\ket{\psi(\vect{\theta})}$, where $\vect{\theta}$ is a vector of parameters that must be iteratively adjusted in order to minimize the expectation value of the Hamiltonian $\bra{\psi(\vect{\theta})}\hat{H}\ket{\psi(\vect{\theta})}$. The Ans\"atz circuit $V(\vect{\theta})$ such that $\ket{\psi(\vect{\theta})} = V(\vect{\theta})\ket{0}$ must be chosen carefully. It is desirable to keep the Ans\"atz circuit as shallow as possible to reduce the effects of noise. Instead of making an arbitrary choice of gates, which is generally suboptimal, it is better to employ an algorithm which learns a good circuit structure at fixed depth to minimize the cost function. For example, the Rotoselect algorithm proposed by Ostaszewski et al. \cite{Ostaszewski2021} is an efficient method for simultaneously optimizing both the structure and parameter values of quantum circuits with only a small computational overhead. The algorithm works by updating the parameters $\vect{\theta}$ and gate choices one at a time according to a closed-form expression for the optimal value of the $d^\text{th}$ parameter $\theta^*_d = \text{argmin}_{\theta_d} \langle H \rangle_{\theta_d}$.

The VQE algorithm varies the Ans\"atz through the use of a parametrized circuit with a fixed form. Such a circuit is often called a \textit{variational form}, and its action may be represented by the linear transformation $V(\vect{\theta})$ such that $\ket{\psi(\vect{\theta}))} = V(\vect{\theta})\ket{0}$. A fixed variational form with a polynomial number of parameters can only generate transformations to a polynomially sized subspace of all the states in an exponentially sized Hilbert space. As a consequence, different variational forms exist. Some of them utilize domain-specific knowledge to generate close approximations, based on the structure of the problem. Others are heuristically designed, not considering the target domain. Once an efficiently parametrized variational form has been chosen, in accordance with the variational method, it is necessary to optimize its parameters to minimize the expectation value of the target Hamiltonian. 

There are sound theoretical arguments for the polynomial scaling of VQE \cite{Peruzzo2014,McClean2016}. However, a number of potential limitations have been identified as well, which could impede the VQE to achieve quantum advantage \cite{Tilly2021}. The key open question is to know whether the variational form can be optimized in a polynomial number of iterations and converge to an approximate yet accurate enough solution.

In QAOA, the depth of the quantum circuit is $p$ times the number of constraints that characterize the optimization problem, where $p \geq 1$ is an integer. Being $2p$ the number of parameters of the quantum circuit, the quality of the approximation improves at the expense of execution time as $p$ is increased \cite{FarhiQuantumSupremacy2019, Nannicini2019}. In the problem we considered, the number of constraints is equal to $M+N+3MF+FK+3L$.
Assuming that the number of flows $F$ is proportional to $N$ and the number of switches $K$ is lower than $M$, then the number of constraints is upper bounded by $O(MN)$. 
For fixed $p$ (with respect to $Q$), and problems with objective function given by a sum of locally acting terms, the optimal angles can be computed in time polynomial in $Q$ \cite{FarhiQuantumApproximate2014}. These considerations, together with the fact that QAOA cannot be efficiently simulated by means of classical computers, make QAOA an appealing approach to explore on forthcoming quantum machines.

\section{Conclusions}
\label{sec:conclusions}

VM consolidation with joint minimization of server and network power consumption is a challenging combinatorial optimization problem. Its equivalent formulation as a QUBO model, provided in this work for the first time, is suitable for being solved by means of quantum algorithms like QAOA which is efficient in terms of space and time complexity (i.e., quantum memory and execution time). Instead, the resource usage of classical solvers grows exponentially (despite the use of B\&B techniques).

Problem instances of interest require more qubits than those provided by current noisy intermediate-scale quantum (NISQ) computers. These devices are composed of hundreds of noisy qubits, i.e., qubits that are not error-corrected, and therefore perform imperfect operations in a limited coherence time. An interesting approach to leverage the limited available resources to perform combinatorial optimization tasks that are classically challenging, has been proposed by Gambella and Simonetto \cite{Gambella2020}. With reference to quadratic-plus-convex mixed binary optimiziation (MBO) problems, the idea is to split the MBO into a binary unconstrained problem that can be solved with quantum algorithms (such as VQE and QAOA), and continuous constrained convex sub-problems that can be solved cheaply with classical algorithms. In a future work, we plan to apply this approach to our binary model, to obtain a reduced QUBO model from the split procedure. In this way, we might solve reasonable problem instances with current NISQ devices.

Although the topic has not been explored in this paper, there are techniques for efficiently mapping inequality constraints in the context of quantum annealers~\cite{Vysko2019}. These techniques may represent an alternative approach to reduce the space complexity of the considered problem. However, quantum annealers are not universal quantum devices, thus preventing the use of combined techniques. 

\section*{Acknowledgment}

This research benefited from the HPC (High Performance Computing) facility of the University of Parma, Italy.

\end{document}